\documentclass[3p,12pt,authoryear]{elsarticle}

\usepackage[utf8]{inputenc}
\usepackage[T1]{fontenc}
\usepackage[round]{natbib}
\usepackage{graphicx}
\usepackage{array}
\usepackage{caption}
\usepackage{float}
\usepackage{booktabs}
\usepackage{adjustbox}

\usepackage{amsmath,amssymb,amsfonts}
\usepackage{graphicx}
\usepackage{adjustbox}     
\usepackage{booktabs}      
\usepackage{array}         
\usepackage{caption}
\usepackage{subcaption}
\usepackage{float}
\usepackage{booktabs}
\usepackage{multirow}
\usepackage{adjustbox}
\usepackage{algorithm}
\usepackage{hyperref}
\usepackage{xcolor}
\usepackage{url}
\usepackage{amsmath}
\usepackage{adjustbox}
\usepackage{booktabs}
\usepackage{array}
\usepackage{float}
\usepackage{algorithm}
\usepackage{algpseudocode}
\usepackage{subcaption}
\usepackage{caption}

\usepackage{lineno}
\modulolinenumbers[5]

\biboptions{authoryear,sort&compress}

\begin{document}

\begin{frontmatter}

\title{\textbf{MOSAIC: A Multi-View 2.5D Organ Slice Selector with Cross-Attentional Reasoning for Anatomically-Aware CT Localization in Medical Organ Segmentation}}
\vspace{2em}
\author{Hania Ghouse}

\author{Muzammil Behzad\corref{cor1}}

\address{King Fahd University of Petroleum and Minerals, Saudi Arabia}
\ead{muzammil.behzad@kfupm.edu.sa}

\vspace{-0.5em}
\cortext[cor1]{Corresponding author}

\begin{abstract}
Efficient and accurate multi-organ segmentation from abdominal CT volumes is a fundamental challenge in medical image analysis. Existing 3D segmentation approaches are computationally and memory intensive, often processing entire volumes that contain many anatomically irrelevant slices. Meanwhile, 2D methods suffer from class imbalance and lack cross-view contextual awareness. To address these limitations, we propose a novel, anatomically-aware slice selector pipeline that reduces input volume prior to segmentation. Our unified framework introduces a vision-language model (VLM) for cross-view organ presence detection using fused tri-slice (2.5D) representations from axial, sagittal, and coronal planes. Our proposed model acts as an "expert" in anatomical localization, reasoning over multi-view representations to selectively retain slices with high structural relevance. This enables spatially consistent filtering across orientations while preserving contextual cues. More importantly, since standard segmentation metrics such as Dice or IoU fail to measure the spatial precision of such slice selection, we introduce a novel metric, Slice Localization Concordance (SLC), which jointly captures anatomical coverage and spatial alignment with organ-centric reference slices. Unlike segmentation-specific metrics, SLC provides a model-agnostic evaluation of localization fidelity. Our model offers substantial improvement gains against several baselines across all organs, demonstrating both accurate and reliable organ-focused slice filtering. These results show that our method enables efficient and spatially consistent organ filtering, thereby significantly reducing downstream segmentation cost while maintaining high anatomical fidelity.

\begin{keyword}
CT scan  \sep 2D segmentation \sep 2.5D segmentation \sep Vision Transformer  \sep  medical image analysis 
\end{keyword}
    \end{abstract}
  \end{frontmatter}

\section{Introduction}
Medical organ segmentation lies at the heart of modern diagnostic and therapeutic workflows, enabling precise identification of anatomical structures that are critical for disease detection, treatment planning, and surgical guidance \citep{Potineni2025}. The unparalleled spatial resolution and cross-sectional imaging capabilities of Computed tomography (CT) scans have made them  indispensable for such tasks \citep{Bhuiyan2023}. However, the computational demands of processing volumetric 3D CT data remain a significant barrier to real-time clinical adoption. While excelling in capturing complex spatial relationships across entire volumes, traditional 3D segmentation methods require substantial computational resources, often rendering them impractical for resource-constrained environments\citep{Dai2022}. For instance, 3D convolutional neural networks (CNNs) achieve superior accuracy by analyzing the full volumetric context but face challenges in deployment due to their high-memory and processing requirements \citep{2025}. At the same time the analysis of medical images, particularly CT scans, is essential for various clinical applications, including computer-aided diagnosis, surgical navigation, and radiation therapy\citep{4}.

\begin{figure}[t!]
\centering
\captionsetup{skip=0.5pt}

\begin{subfigure}[b]{0.32\textwidth}
  \centering
  \includegraphics[height=3.8cm]{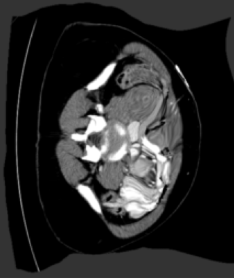}
  \caption{Axial Slice (Horizontal View)}
\end{subfigure}
\hfill
\begin{subfigure}[b]{0.32\textwidth}
  \centering
  \includegraphics[height=3.8cm]{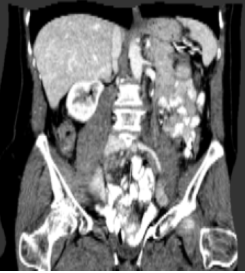}
  \caption{Coronal Slice (Frontal View)}
\end{subfigure}
\hfill
\begin{subfigure}[b]{0.32\textwidth}
  \centering
  \includegraphics[height=3.8cm]{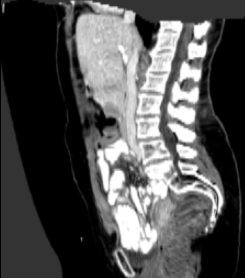}
  \caption{Sagittal Slice (Side View)}
\end{subfigure}

\vspace{0.2em}
\caption{Orthogonal anatomical views used in CT imaging.}
\label{fig:ct-views}
\end{figure}

In contrast, the conventional 2D approaches in the literature process individual slices sequentially and offer computational efficiency but at the cost of losing inter-slice spatial context \citep{Zhou2023}. This limitation is particularly problematic in  multi organ segmentation, where critical anatomical details often span multiple slices \citep{5}. Clinical practitioners and radiologists intuitively mitigate this bottleneck by leveraging three orthogonal views: axial, sagittal, and coronal to triangulate anatomical boundaries and assess the spatial relationships between structures \citep{6}, as illustrated in Figure \ref{fig:ct-views}. 

\begin{figure}[t!]
    \centering
    \includegraphics[width=0.6\textwidth]{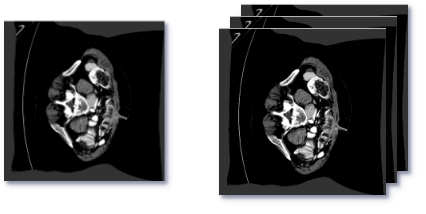}
    \caption{Illustration of input representations in 2D (left side) and 2.5D (right side)  slice.}
    \label{fig:segmentation_inputs}
\end{figure}

Despite the context-rich spatial information, most automated segmentation studies focus exclusively on axial slices to utilize conventional 2D deep learning models, neglecting sagittal and coronal views that could provide complementary spatial information \citep{Lindeijer}. This narrow focus magnifies the inherent weaknesses of 2D methods, such as their inability to capture 3D structural continuity, leading to suboptimal segmentation accuracy in regions with complex geometries \citep{Yu}. Furthermore, peripheral slices often contain irrelevant structures like muscle, bone, or artifacts, while central slices may lack sufficient anatomical details to guide segmentation \citep{7}. Existing 2D frameworks typically process all slices uniformly, wasting computational resources on unwanted regions and failing to prioritize slices with high organ presence \citep{8}. This inefficiency is intensified by class imbalancing, as datasets often contain far more non-organ slices than those with target structures, thereby skewing the model performance and hindering the training of robust organ segmentation algorithms\citep{Tappeiner2022}. To bridge the gap between 2D efficiency and 3D accuracy, advances in 2.5D segmentation aim to integrate adjacent slices (e.g., current, previous, and next), to mimic volumetric context \citep{9}, \citep{10}, \citep{11},as shown in Figure ~\ref{fig:segmentation_inputs}. While this approach reduces computational overhead compared to full 3D analysis, it still processes slices in isolation and neglects the unique anatomical insights provided by multi-plane views. For example, the sagittal view may better capture longitudinal organ contours, while the coronal view highlights lateral structures. This crucial information is lost when focusing solely on axial slices \citep{Khan2019}.

\section{Literature Review}
Organ segmentation in CT scans has advanced significantly with recent advancements in deep learning, yet several challenges persist in balancing computational efficiency, anatomical specificity, and generalizability\citep{Ni2024} \citep{10092740}. Traditional approaches, like deformable models \citep{kumar2022deformable} \citep{Kim2022} and graph cuts \citep{12} \citep{Xie2025} \citep{graph}, rely on shape priors and energy minimization to segment organs, but these methods often struggle with inter-patient anatomical variability and require manual initialization, limiting their scalability. Other deep learning efforts, including U-Net variants\citep{Brunese2025} \citep{shahstry}, demonstrated promising performance in 2D segmentation but faced class imbalance issues due to non-uniform slice relevance across CT volumes.For example, a two-step framework that integrates a slice selector with 3D encoders achieved a Dice score of 95\% for organs at risk in head and neck cancer. However, it still processed redundant slices, resulting in increased computational overhead \citep{13},\citep{Guo2020},\citep{Chen2021},\citep{Lei2021}. Recent 3D models, such as V-Net and 3D U-Net  \citep{14} \citep{Pan2025} \citep{Hossain2025}, capture spatial context more effectively but demand high computational resources, hindering their adoption in clinics with limited infrastructure. To address these limitations, researchers have explored 2.5D methods that incorporate neighboring slices to approximate the volumetric context without the full computational cost of 3D models \citep{15} \citep{Yin2025}. 

One approach involves thickening 2D network inputs by feeding multiple slices as channels to integrate 3D contextual information while maintaining lower inference latency \citep{16} \citep{Vu2020}. Other studies used a 2.5D residual U-Net for liver lesion segmentation and demonstrated improved performance over 2D models.However, they still lack comprehensive inter-slice spatial relationships \citep{17} \citep{Zhao2025}.Similarly, attention mechanisms have also been applied to focus on relevant features, with volumetric attention models outperforming 2D counterparts in multi-organ segmentation \citep{18} \citep{zheng2024}. Despite these advancements, the comparative performance of 2.5D versus 3D methods remains unclear, as evidenced by empirical studies on tasks involving different modalities and targets \citep{Zhang2022}. Meanwhile, cascaded architectures like the multi-dimensional cascaded net (MDCNet) \citep{21} addressed class imbalance by hierarchically segmenting organs, but their layered design introduced latency in processing high-resolution CT volumes.  Recent advances in multi-organ segmentation have leveraged ensemble methods to improve robustness. A study \citep{20} proposed an ensemble of 14 datasets to segment organs and tumors using vision-language models (VLMs) based on the contrastive language-image pre-training (CLIP) model \citep{pmlr-v139-radford21a}, achieving 91\% accuracy across 3,410 CT scans. Building upon this, MedCLIP-SAMv2 \citep{medclip} integrates CLIP with the Segment Anything Model (SAM) \citep{sam} to perform segmentation on clinical scans using text prompts in both zero-shot and weakly supervised settings. Furthermore, SaLIP  \citep{salip} presents a unified framework combining SAM and CLIP for zero-shot medical image segmentation. These studies collectively highlight the potential of VLM-based models in advancing medical image segmentation and classification tasks, while also acknowledging the challenges related to computational efficiency and the need for further exploration in real-time applications. 
\subsection{Challenges and Limitations}

Existing 2.5D methods, such as slice-wise regression \citep{22} or axial-plane CNNs \citep{23}, prioritize efficiency but sacrifice spatial context across sagittal and coronal views, leading to suboptimal delineation of organs like the liver or kidneys, where longitudinal and lateral contours are critical. Most studies focus on axial slices, neglecting sagittal and coronal views that radiologists routinely use to resolve ambiguities \citep{24}. Consequently, current models either overprocess non-informative slices or underutilize anatomical cues, resulting in trade-offs between accuracy and computational cost. This gap motivates our 2.5D multi-plane framework, which selectively prioritizes slices and integrates orthogonal views to enhance efficiency without compromising diagnostic rigor.

\subsection{Contributions}
To address the limitations of computational inefficiency and anatomical inconsistency in current CT processing methods, we propose MOSAIC: a Multi-view Organ Slice-selector for Anatomically Informed CT localization model, that imitate radiologist-style reasoning by selectively retaining only the most informative slices across axial, coronal, and sagittal views. The pipeline begins with a lightweight slice filtering module that discards structurally irrelevant slices, thereby reducing redundant computation while preserving anatomical coverage. Following this, the remaining slices are processed using a multi-view integration module that fuses orthogonal representations through fine-tuning the CLIP as our model's backbone. This enables our model to align local slice features with organ-specific prompts and reason jointly across views, improving semantic precision and spatial coherence. To better evaluate the effectiveness of this slice selection strategy, we introduce Slice Localization Concordance (SLC), a novel metric that quantifies both anatomical coverage and spatial alignment with ground-truth organ locations. Unlike traditional performance scores such as F1 or PR-AUC, our new metric SLC directly captures the localization fidelity of the selected slices and offer a more clinically meaningful measure of performance. Overall, MOSAIC provides an efficient, anatomically aware, and interpretable approach to pre-segmentation volume reduction in abdominal CT for effective medical image analysis.

\section{Methodology}

\subsection{Data Pre-processing and Multi-View Slice Extraction}
\begin{figure}[t!]
    \centering
    \includegraphics[width=0.8\linewidth]{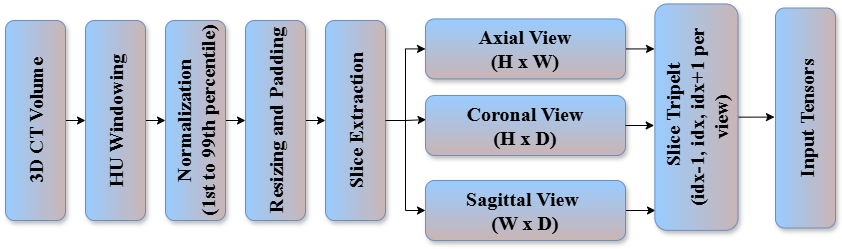}
    \caption{Multi-View Slice Extraction Pipeline.}
    \label{fig:flowchart}
\end{figure}

To prepare volumetric CT data for organ presence slice selection, a sequence of standardized preprocessing operations is applied. As shown in the complete pipeline in Figure~\ref{fig:flowchart}. Each 3D CT volume \( V \in \mathbb{R}^{H \times W \times D} \) is first converted from hounsfield units (HU) to normalized intensity values. This begins with HU windowing in the soft tissue range of \([-50, 200]\), where all voxel intensities are clipped and linearly mapped within this interval.
Following windowing, we apply a percentile-based normalization to reduce inter-subject intensity variability and suppress outlier values. Given an input volume \( V \), the normalized intensity at voxel \( x \in V \) is computed as:

\begin{equation}
\tilde{V}(x) = \frac{V(x) - P_1(V)}{P_{99}(V) - P_1(V)},
\label{eq:percentile-norm}
\end{equation}where \( P_1(V) \) and \( P_{99}(V) \) denote the 1st and 99th percentile intensities of the volume \( V \), respectively, and \( \tilde{V}(x) \in [0, 1] \) represents the normalized value. This normalization step ensures that the intensity distribution is standardized across subjects while preserving contrast in relevant anatomical regions. Each extracted slice is subsequently resized and zero-padded to a fixed resolution of \(256 \times 256\) pixels, preserving aspect ratio and spatial fidelity. After the slices are spatially adjusted the three orthogonal views are extracted from the normalized volume to capture complementary anatomical information. Axial slices are indexed along the depth dimensions as:
\begin{equation}
S_{\text{axial}}(z) = V[:, :, z], \quad z \in [0, D{-}1].
\label{eq:axial}
\end{equation}
Similarly, coronal slices are extracted along the width axis and rotated by \(90^\circ\) to maintain consistent anatomical orientation as:
\begin{equation}
S_{\text{coronal}}(y) = \mathcal{R}_{90}\big(V(:, y, :)\big), \quad y \in \{0, 1, \dots, W{-}1\},
\label{eq:coronal}
\end{equation}where \( \mathcal{R}_{90} \) denotes a counter-clockwise in-plane rotation by \(90^\circ\).
Lastly sagittal slices are obtained along the height axis and transposed prior to rotation as:
\begin{equation}
S_{\text{sagittal}}(x) = \mathcal{R}_{90}\big(V(x, :, :)^\top\big), \quad x \in \{0, 1, \dots, H{-}1\}.
\label{eq:sagittal}
\end{equation}
To encode local spatial context, each selected slice \( S_i \in \mathbb{R}^{H \times W} \) is expanded into an n-channel tensor by stacking it with its immediate neighbors along the channel axis. The resulting contextualized slice representation \( \mathbf{I}_i \in \mathbb{R}^{n \times H \times W} \) is defined as:

\begin{equation}
\mathbf{I}_i =
\begin{cases}
\left[\, S_{i-1},\; S_i,\; S_{i+1} \,\right], & \text{if } 1 \leq i \leq N{-}2, \\
\left[\, S_i,\; S_i,\; S_i \,\right], & \text{otherwise},
\end{cases}
\label{eq:triplet-stack}
\end{equation}where \( N \) denotes the total number of selected slices in a given anatomical plane, and \( [\,\cdot\,] \) indicates channel-wise concatenation.To construct a unified multi-view input, the triplet tensors from the three orthogonal views are concatenated into a single tensor \( \mathbf{I} \in \mathbb{R}^{m \times H \times W} \) as follows:

\begin{equation}
\mathbf{I} = \left[\, \mathbf{I}^{(a)},\; \mathbf{I}^{(c)},\; \mathbf{I}^{(s)} \,\right],
\label{eq:tri-view-input}
\end{equation}where \( \mathbf{I}^{(a)} \), \( \mathbf{I}^{(c)} \), and \( \mathbf{I}^{(s)} \) correspond to the contextualized triplets from the axial, coronal, and sagittal views, respectively.To improve robustness during training, we apply spatial augmentations to each input tensor, including random flips along the horizontal and vertical axes, as well as in-plane rotations sampled from a bounded angle range. This pipeline ensures that each input volume is anatomically coherent, contextually enriched, and spatially standardized across orientations.

\subsection{Model Framework}

Our proposed MOSAIC framework is a multi-stage pipeline designed to reduce redundant CT slices while preserving organ-relevant anatomical content. It decomposes full volumes into 2D slices, filters non-informative views, constructs a fused 2.5D multi-view input, and performs organ-presence prediction using a fine-tuned model based on CLIP. The system consists of three main components: (1) a lightweight slice filtering module, (2) multi-view 2.5D input construction with cross-view attention fusion, and (3) a multimodal prediction module based on vision-language semantic information extraction.

\subsubsection{Slice Filtering Module}

\begin{figure}[t!]
\centering
\includegraphics[width=\linewidth]{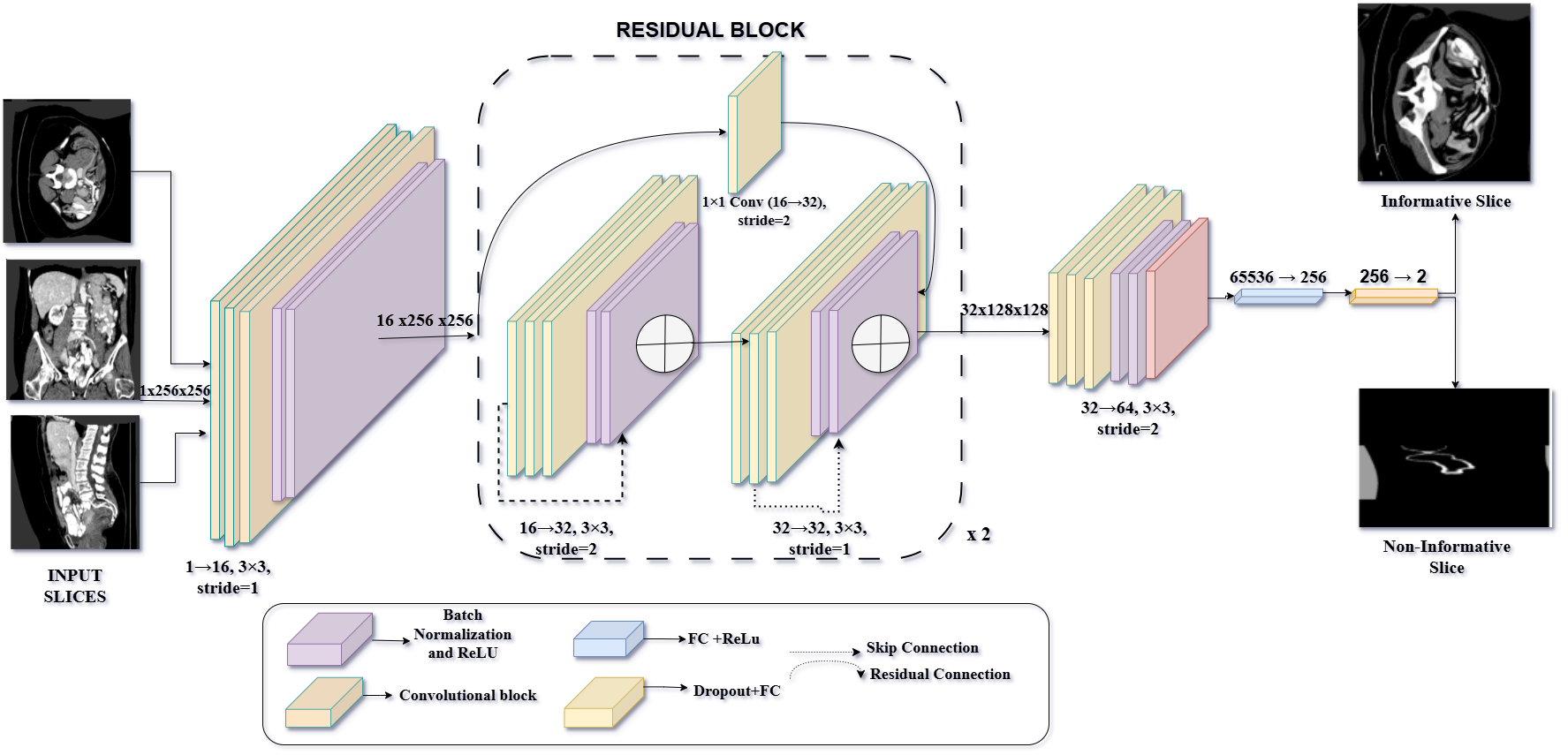}
\caption{Overview of the proposed Stage 1 Filtering Module.}
\label{fig:stage1}
\end{figure}

To eliminate redundant computation and focus on informative anatomy, we propose a slice-level filtering module that removes non-relevant slices across axial, coronal, and sagittal views. As clearly observed in CT scans, many CT slices contain no meaningful foreground content, which dilutes training and inference. Given a CT scan \( V \in \mathbb{R}^{H \times W \times D} \) and its segmentation map \( L \in \mathbb{Z}^{H \times W \times D} \), the volume is sliced along each plane. Each slice \( x_s^{(v)} \in \mathbb{R}^{H \times W} \), from view \( v \in \{x, y, z\} \), is assigned a binary informativeness label based on its foreground pixel ratio:

\begin{equation}
y_s^{(v)} =
\begin{cases}
1, & \text{if } \displaystyle \sum_{i,j} \mathbb{1}\big(L_s^{(v)}(i,j) > 0\big) \geq \tau \cdot H \cdot W \\
0, & \text{otherwise}.
\end{cases}
\label{eq:filter-thresh}
\end{equation}A shallow CNN with two residual blocks and a \(1 \times 1\) bottleneck convolution is trained to predict this binary label, as shown in Figure \ref{fig:stage1}. To handle the imbalance between informative and non-informative slices, we employ class-weighted binary cross-entropy and enforce volume-level data splits to prevent leakage.

\subsubsection{Multi-View Feature Fusion via Cross Attention}

Each retained slice is converted into a triplet by stacking it with its adjacent slices forming tensors \( \mathbf{I}^{(v)}_s \in \mathbb{R}^{n \times H \times W} \). The triplets of all three planes are then concatenated channel-wise to form a unified m-channel input tensor \( \mathbf{I} \in \mathbb{R}^{m \times H \times W} \). The selected architecture independently encodes features from each plane, producing outputs \( f^{(x)}, f^{(y)}, f^{(z)} \in \mathbb{R}^{C \times H' \times W'} \) which are  then flattened and concatenated into a unified token sequence:
\begin{equation}
\mathbf{F} = \left[ \text{vec}(f^{(x)}),\; \text{vec}(f^{(y)}),\; \text{vec}(f^{(z)}) \right] \in \mathbb{R}^{nH'W' \times C}.
\end{equation}Multi-head attention is applied over the token sequence with projections \( Q = \mathbf{F}W_Q,\; K = \mathbf{F}W_K,\; V = \mathbf{F}W_V \in \mathbb{R}^{nH'W' \times d} \), where \( W_Q, W_K, W_V \in \mathbb{R}^{C \times d} \). The attention output is given by:
\begin{equation}
\widetilde{\widetilde{\mathcal{A}}}(\mathbf{Q}, \mathbf{K}, \mathbf{V}) = \text{softmax}\left( \frac{\mathbf{Q} \mathbf{K}^\top}{\sqrt{d}} \right) \mathbf{V}.
\end{equation}The fused output is computed via a residual connection followed by layer normalization and a feed-forward transformation:
\begin{equation}
\mathbf{Z} = \mathcal{W}_2 \, \phi\left( \mathcal{W}_1 \, \mathcal{N} \left( \mathbf{F} + \widetilde{\widetilde{\mathcal{A}}}(\mathbf{Q}, \mathbf{K}, \mathbf{V}) \right) \right),
\end{equation}where \( \mathcal{N} \) denotes layer normalization, \( \mathcal{W}_1 \in \mathbb{R}^{C \times C'} \), \( \mathcal{W}_2 \in \mathbb{R}^{C' \times C} \) are learnable weight matrices of the multi-layer perceptron, and \( \phi \) refers to a pointwise non-linear activation function (e.g., ReLU).

\begin{figure}[t!]
\centering
\includegraphics[width=\linewidth]{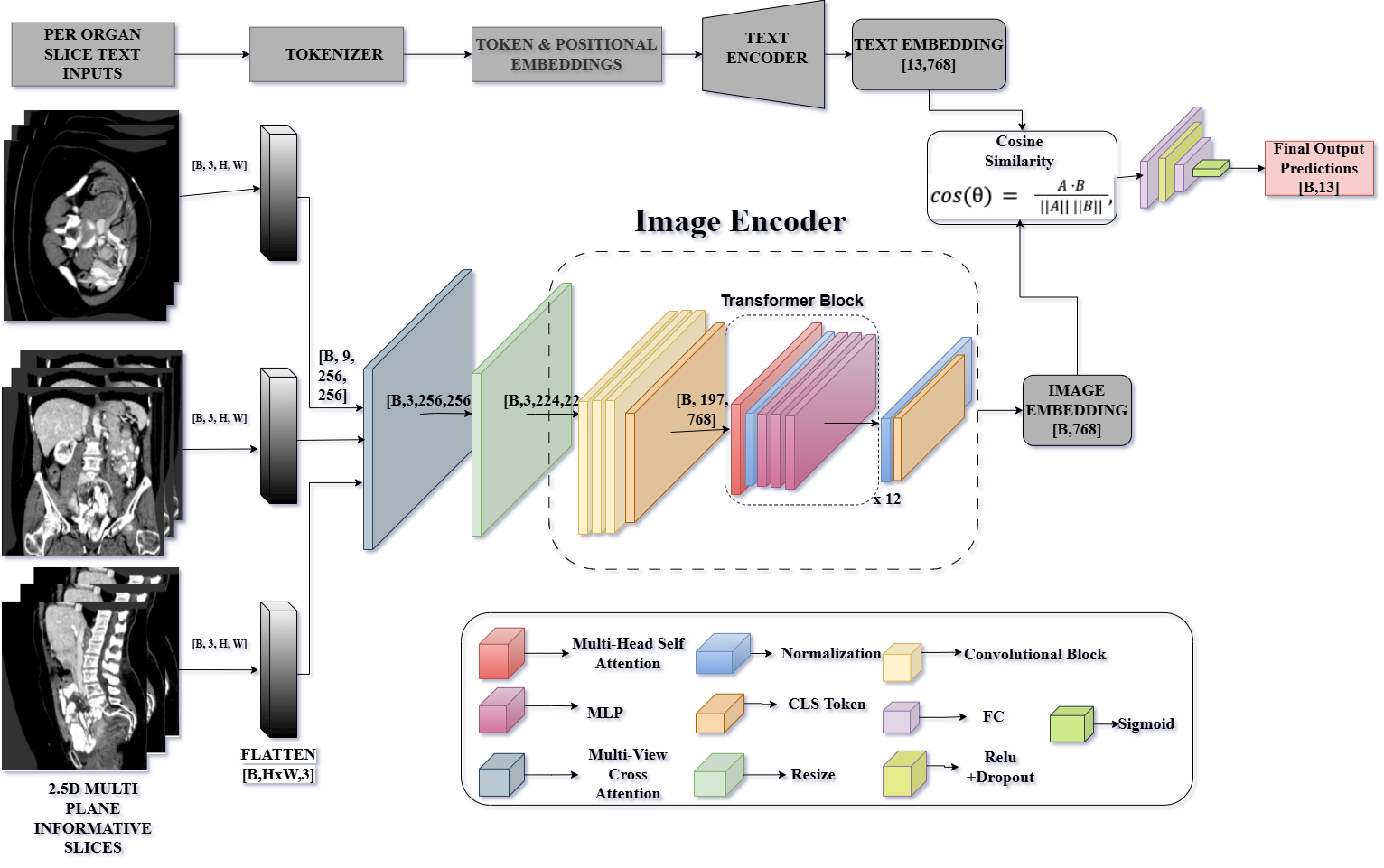}
\caption{Overview of the proposed Multi-View 2.5D slice selector framework.}
\label{fig:framework}
\end{figure}
\subsubsection{Multimodal Organ Slice Selector with Vision-Language Encoder}

As illustrated in Figure \ref{fig:framework}, the fused token representation \( \mathbf{Z} \) is reshaped into a n-channel image tensor, passed through a \(1 \times 1\) convolution to reduce channel dimensions, and resized to \(224 \times 224\). The visual encoder then computes an image embedding \( \mathbf{z}_{\text{img}} \in \mathbb{R}^{D} \) (e.g., \(D=768\)).

\subsubsection{Prompt Construction for Vision-Language-Based Slice Selection}

We simultaneously generate a bank of synthetic prompts using a structured template for each organ.To provide organ-specific guidance, we construct textual prompts of the form:

\[
\texttt{``a CT \textless noun\textgreater\ \textless verb\textgreater\ the \textless organ\textgreater''}
\]

from the following vocabularies:
\[
\mathcal{S} = \{\text{slice},\, \text{scan},\, \text{image},\, \text{view}\}, \qquad
\mathcal{V} = \{\text{showing},\, \text{depicting},\, \text{highlighting},\, \text{containing}\}
\]For each organ \( o_k \in \mathcal{O} \), we generate 16 prompt variations (4 nouns × 4 verbs). These prompts are encoded via the text encoder into normalized embeddings \( \mathbf{z}_{\text{text}}^{(k)} \in \mathbb{R}^{D} \), where \( k = 1, \dots, N \) and \( N \) is the number of organ classes.

\begin{table}[b!]
  \centering
  \caption{Sample synthetic prompts generated per organ.}
  \label{tab:prompt-examples}
  \begin{tabular}{@{}ll@{}}
    \toprule
    \textbf{Organ} & \textbf{Sample Prompts} \\
    \midrule
    Spleen
    & \texttt{a CT slice showing the spleen} \\
    & \texttt{a CT scan depicting the spleen} \\
    & \texttt{a CT image highlighting the spleen} \\
    & \(\dots\) \\
    \addlinespace
    Liver
    & \texttt{a CT view showing the liver} \\
    & \texttt{a CT image containing the liver} \\
    & \texttt{a CT scan highlighting the liver} \\
    & \(\dots\) \\
    \bottomrule
  \end{tabular}
\end{table}

\subsubsection{Multimodal Alignment and Prediction}

In order to align the multimodal embeddings, we compute the cosine similarities between the image embedding and each text embedding:

\begin{equation}
s_k = \cos\left(\mathbf{z}_{\text{img}}, \mathbf{z}_{\text{text}}^{(k)}\right) = \frac{\mathbf{z}_{\text{img}} \cdot \mathbf{z}_{\text{text}}^{(k)}}{\|\mathbf{z}_{\text{img}}\| \, \|\mathbf{z}_{\text{text}}^{(k)}\|}, \quad k = 1, \dots, N.    
\end{equation}The similarity vector \( \mathbf{s} \in \mathbb{R}^{N} \) is concatenated with \( \mathbf{z}_{\text{img}} \), forming a joint embedding \( [\mathbf{z}_{\text{img}};\, \mathbf{s}] \in \mathbb{R}^{D+N} \). This is then fed through an MLP with sigmoid activationto output:

\begin{equation}
\hat{\mathbf{y}} = \text{MLP}\left([\mathbf{z}_{\text{img}};\, \mathbf{s}]\right) \in [0,1]^N    
\end{equation}

\subsubsection{Loss Function}

We adopt Class-Balanced Focal Loss (CBFL) to mitigate class imbalance:

\begin{equation}
\mathcal{L}_{\text{CBFL}} = \sum_{k=1}^{N} \alpha_k (1 - \hat{y}_k)^\gamma y_k \log(\hat{y}_k) + (1 - \alpha_k) \hat{y}_k^\gamma (1 - y_k) \log(1 - \hat{y}_k),
\end{equation}where \( y_k \in \{0,1\} \) is the binary ground truth, \( \alpha_k \) is the inverse frequency of class \( k \), \( \hat{y}_k \in [0,1] \) is the predicted probability, and \( \gamma = 2 \) is the focusing parameter. This multimodal formulation allows MOSAIC to integrate spatial context, anatomical semantics, and language supervision for efficient and anatomically-aware slice filtering prior to segmentation.

\subsubsection{Slice Localization Concordance (SLC)}

The volumetric CT scans often contain organs that span multiple slices, yet most slice-level prediction methods operate on isolated 2D inputs ignoring spatial continuity and anatomical context. While conventional metrics such as F1-score or PR-AUC are widely used for evaluating slice-level predictions, they offer limited insight into the spatial fidelity of predictions particularly in terms of proximity to the true anatomical center. This is especially problematic in clinical tasks such as organ localization and surgical planning, where even semantically correct predictions may be clinically irrelevant if they are spatially misaligned. To address this, we introduce a new evaluation metric: Slice Localization Concordance (SLC), which jointly assesses anatomical coverage and spatial proximity, thereby offering a more clinically meaningful measure of spatial accuracy.

Let \( A_s^{(o)} \) denote the number of pixels labeled as organ \( o \) in slice \( s \), and let the anatomical center slice \( s^* \) be the one with maximum organ area:

\begin{equation}
A_o^* = \max_s A_s^{(o)}, \qquad s^* = \arg\max_s A_s^{(o)}.
\label{eq:slc-center}
\end{equation}Given a set of predicted slices \( S_o \subset \{0, 1, \dots, N{-}1\} \) for organ \( o \), we define a normalized coverage score \( C_s \) and a proximity weight \( w_s \) for each slice \( s \in S_o \). The coverage score quantifies the anatomical relevance of a predicted slice relative to the true center:

\begin{equation}
C_s = \frac{A_s^{(o)}}{A_o^* + \varepsilon},
\label{eq:slc-coverage}
\end{equation}
where \( \varepsilon \) is a small constant to avoid division by zero. The proximity weight emphasizes predictions closer to the anatomical center:

\begin{equation}
w_s = \exp\left( -\frac{|s - s^*|}{\delta} \right),
\label{eq:slc-weight}
\end{equation}where \( \delta > 0 \) is a tunable tolerance parameter controlling the spatial decay. The final SLC score for organ \( o \) is computed as the proximity-weighted average of coverage scores:

\begin{equation}
\text{SLC}_o = 
\begin{cases}
\displaystyle \frac{ \sum_{s \in S_o} C_s w_s }{ \sum_{s \in S_o} w_s }, & \text{if } S_o \neq \emptyset,  \\
0, & \text{if } S_o = \emptyset,
\end{cases}
\label{eq:slc-final}
\end{equation}where slices with higher anatomical content and smaller deviation from \( s^* \) contribute more to the final score. In the case where organ \( o \) is absent from the ground truth volume (i.e., \( A_s^{(o)} = 0 \) for all \( s \)), the SLC score is zero and excluded from evaluation. This formulation bridges semantic correctness with anatomical alignment by directly incorporating both organ visibility and proximity into the scoring function. It penalizes distant but technically correct predictions and rewards anatomically meaningful selections. The overall computational flow of SLC is shown in Algorithm \ref{alg:slc}. The metric is particularly suitable for evaluating 2.5D and multi-view models like MOSAIC, where the goal is not merely slice-level accuracy but accurate spatial localization across views.

\begin{algorithm}[t!]
\caption{Computation of Slice Localization Concordance (SLC)}
\label{alg:slc}
\begin{algorithmic}[1]
\Require Organ-wise pixel counts \( A_s^{(o,v)} \), predicted slices \( S_o^{(v)} \), tolerance \( \delta \)
\Ensure SLC scores \( \text{SLC}_o^{(v)} \) for all organs \( o \) and views \( v \in \{x, y, z\} \)
\ForAll{organs \( o \)}
  \ForAll{views \( v \in \{x, y, z\} \)}
    \State \( A^* \gets \max_s A_s^{(o,v)} \), \quad \( s^* \gets \arg\max_s A_s^{(o,v)} \)
    \State \( S \gets S_o^{(v)} \); \textbf{if} \( S = \emptyset \) \textbf{then} \( \text{SLC}_o^{(v)} \gets 0 \); \textbf{continue}
    \State \( N \gets 0 \), \quad \( D \gets 0 \)
    \ForAll{\( s \in S \)}
      \State \( C_s \gets \frac{A_s^{(o,v)}}{A^* + \varepsilon} \), \quad \( w_s \gets \exp\left(-\frac{|s - s^*|}{\delta}\right) \)
      \State \( N \gets N + C_s \cdot w_s \), \quad \( D \gets D + w_s \)
    \EndFor
    \State \( \text{SLC}_o^{(v)} \gets \frac{N}{D} \)
  \EndFor
\EndFor
\State \Return all \( \text{SLC}_o^{(v)} \)
\end{algorithmic}
\end{algorithm}

\section{Experimental Setup}
To validate the effectiveness of our proposed pipeline, we conduct comprehensive experiments focusing on its core components.
\subsection{Dataset}
We adopt the Beyond the Cranial Vault (BTCV) dataset, which is widely recognized benchmark in abdominal organ segmentation, owing to its anatomical diversity and clinically relevant complexity. The dataset comprises 50 contrast-enhanced abdominal CT scans (in-plane resolution: 512 × 512, variable axial depth), each annotated with 13 abdominal organs, including large structures like liver and spleen and small, difficult-to-detect stuctures like  adrenal glands and gallbladder. BTCV is particularly well-suited for evaluating multi-organ segmentation and classification models due to its inclusion of both high-contrast and low-volume organs under realistic imaging scenarios. We show in Table \ref{tab:btcv-stats} that the dataset presents substantial inter-organ imbalance: large-volume organs like the liver and spleen occupy 3.7\% and 3.3\% of the scan on average, whereas structures such as the adrenal glands and gallbladder are less than 0.05\%, highlighting the extreme skew in anatomical representation. Moreover, certain organs are inconsistently annotated across scans. For instance, the gallbladder appears in only 28 of 30 cases that further complicate organ-level learning. This variability introduces meaningful challenges for multi-label slice selector models and justifies the need for class imbalance-aware training strategies, such as focal loss and organ-specific thresholds, as employed in our pipeline.
\renewcommand{\arraystretch}{0.95}
\begin{table}[t!]
\centering
\scriptsize
\caption{Organ-wise Volume Fraction (VF) and Actual Size Statistics in BTCV Dataset}
\label{tab:btcv-stats}
\begin{adjustbox}{max width=\linewidth}
\begin{tabular}{
    l
    r r r r
    r r r r
    r
}
\toprule
\textbf{Organ} & 
\textbf{VF Min} & 
\textbf{VF Med} & 
\textbf{VF Mean} & 
\textbf{VF Max} & 
\textbf{Size Min} & 
\textbf{Size Med} & 
\textbf{Size Mean} & 
\textbf{Size Max} & 
\textbf{Count} \\
 & (\%) & (\%) & (\%) & (\%) & (cm\textsuperscript{3}) & (cm\textsuperscript{3}) & (cm\textsuperscript{3}) & (cm\textsuperscript{3}) &  (\#) \\
\midrule
Whole Foreground & 2.76 & 4.26 & 4.46 & 7.48 & 1766 & 3122 & 3223 & 6963 & 30 \\
Spleen           & 0.13 & 0.44 & 0.32 & 2.76 & 63   & 224  & 338  & 3087 & 30 \\
Right Kidney     & 0.02 & 0.21 & 0.22 & 0.39 & 10   & 151  & 158  & 259  & 30 \\
Left Kidney      & 0.02 & 0.21 & 0.22 & 0.34 & 11   & 161  & 158  & 262  & 30 \\
Gallbladder      & 0.01 & 0.03 & 0.04 & 0.11 & 2    & 17   & 28   & 82   & 28 \\
Esophagus        & 0.01 & 0.02 & 0.02 & 0.05 & 3    & 14   & 16   & 51   & 30 \\
Liver            & 1.58 & 2.45 & 3.70 & 10.33 & 1033 & 1659 & 1738 & 2563 & 30 \\
Stomach          & 0.23 & 0.53 & 0.61 & 1.40 & 139  & 429  & 466  & 1277 & 30 \\
Aorta            & 0.06 & 0.12 & 0.14 & 0.30 & 29   & 85   & 101  & 283  & 30 \\
Inf. Vena Cava   & 0.06 & 0.11 & 0.13 & 0.21 & 27   & 83   & 88   & 175  & 30 \\
Portal/Spl. Vein & 0.03 & 0.04 & 0.05 & 0.13 & 15   & 31   & 35   & 125  & 30 \\
Pancreas         & 0.06 & 0.11 & 0.12 & 0.22 & 41   & 85   & 82   & 137  & 30 \\
Right Adrenal    & 0.00 & 0.01 & 0.01 & 0.01 & 1    & 4    & 4    & 8    & 30 \\
Left Adrenal     & 0.00 & 0.01 & 0.01 & 0.01 & 2    & 5    & 5    & 11   & 30 \\
\bottomrule
\end{tabular}
\end{adjustbox}
\end{table}

\subsection{Implementation Details}
We summarize the experimental setup and training configuration in Table~\ref{tab:training-config}. Our pipeline was trained and evaluated on over 68,000 CT slices spanning multiple anatomical views, ensuring a diverse and representative sample distribution across train, validation, and test splits.

\begin{table}[b!]
\centering
\caption{Training Configuration Summary}
\label{tab:training-config}
\begin{tabular}{ll}
\hline
\textbf{Parameter} & \textbf{Value} \\
\hline
GPU & NVIDIA RTX 3090 (24 GB) \\
Framework & PyTorch 2.0 \\
Batch size & 16 \\
Optimizer & AdamW \\
Initial learning rate & $1 \times 10^{-4}$ \\
LR scheduler & OneCycleLR \\
Epochs & 50 (with early stopping) \\
Input size & $256 \times 256$ \\
Total Train Slices & 42020\\
Total Validation Slices & 11662\\
Total Test Slices & 14004\\
\hline
\end{tabular}
\end{table}

\subsection{Ablation Studies}
To quantify the contribution of individual components, we conduct targeted ablation experiments. Each variant modifies a single design choice while keeping all others fixed. We report average F1, PR-AUC, and SLC across all organs.

\begin{table}[t!]
\centering
\scriptsize
\caption{Model-wise comparison of  performance with standard metrics and our new  slice localization concordance (SLC)}
\label{tab:model_performance_slc}
\begin{tabular}{lcccccc}
\toprule
\textbf{Model} & \textbf{Precision} & \textbf{Recall} & \textbf{F1-Score} & \textbf{ROC-AUC} & \textbf{PR-AUC} & \textbf{Overall SLC} \\
\midrule
EfficientNet-V2s       & 0.9284 & 0.9422 & 0.9356 & 0.9935 & 0.9922 & 0.8821 \\
ResNet50               & 0.9225 & 0.9387 & 0.9291 & 0.9927 & 0.9913 & 0.8548 \\
Swin-T                 & 0.9181 & 0.9443 & 0.9302 & 0.9931 & 0.9918 & 0.8715 \\
EfficientNet-B0        & 0.9091 & 0.9318 & 0.9190 & 0.9908 & 0.9896 & 0.8288 \\
MobileNet              & 0.8940 & 0.9251 & 0.9093 & 0.9883 & 0.9871 & 0.8465 \\
\textbf{MOSAIC (Ours)}  & \textbf{0.9348} & \textbf{0.9497} & \textbf{0.9426} & \textbf{0.9945} & \textbf{0.9934} & \textbf{0.956} \\
\bottomrule
\end{tabular}
\end{table}
\begin{figure}[b!]
    \centering
    \includegraphics[width=0.94\linewidth]{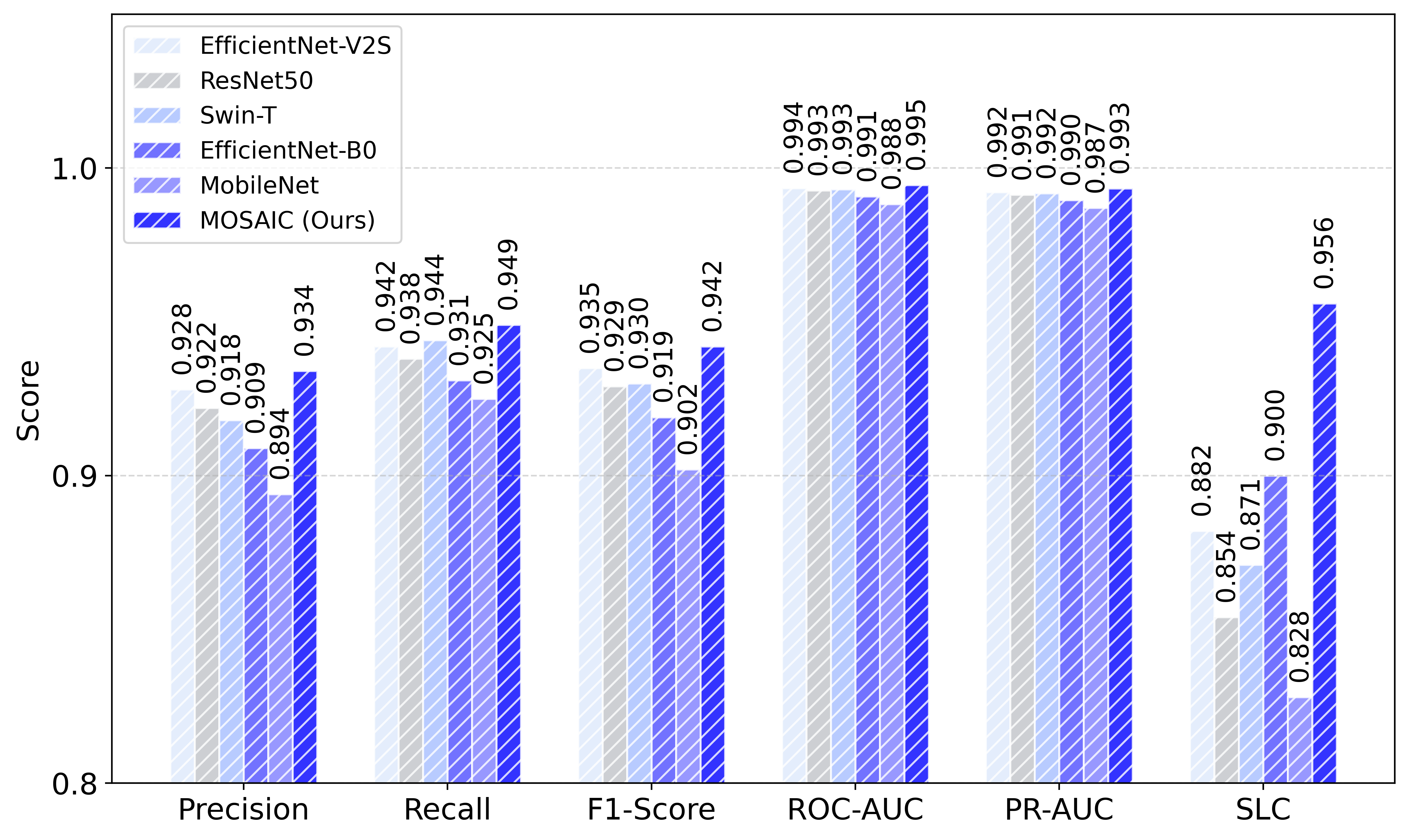}
    \caption{Model-wise Comparisons with standard metrics and our SLC metric}
    \label{fig:overall_evaluation}
\end{figure}
\subsubsection{Baseline Comparison}
In Table \ref{tab:model_performance_slc}, we demonstrate that MOSAIC outperforms all baseline models in both classification and slice-selection tasks. It achieves the highest Precision of 0.934, Recall of 0.949, and F1-Score of 0.942, as well as significant ROC-AUC of 0.9945 and PR-AUC of 0.9934. These results confirm MOSAIC’s strong ability to detect organ presence. Additionally, it records an SLC of 0.956, which is well above the next best model, EfficientNet-V2s, at 0.882. This highlights MOSAIC’s superior anatomical alignment when selecting diagnostically relevant slices.

In Figure \ref{fig:overall_evaluation}, we visualize these six evaluation metrics using bar charts shaded by performance. MOSAIC consistently appears with the darkest bars, indicating its top performance across all measures. EfficientNet-V2s, ResNet-50, and Swin-T show tightly clustered bars in the 0.92–0.99 range for ROC-AUC and PR-AUC, reflecting competitive performance. In contrast, EfficientNet-B0 and MobileNet show lighter bars, illustrating a trade-off between model complexity and accuracy. The figure reinforces MOSAIC’s effectiveness in both detection and anatomical slice selection.

\subsubsection{Evaluating Single vs. Multi-View Representations}
In Table~\ref{tab:clip_ablation_views}, we evaluate the effect of both view type and input dimensionality in Vision-Language model(VLM) based organ presence slice selectors. When limited to axial-only slices, the 2D single-view variant achieves a baseline F1-score of 0.914. By introducing multi-view inputs (axial, coronal, sagittal) in 2D improves performance to 0.921, confirming the advantage of incorporating cross-sectional anatomical context.Further gains are observed when upgrading from 2D to 2.5D inputs. The 2.5D single-view model achieves an F1 of 0.926, while our proposed  Multi View (2.5D) model outperforms all others with a F1-score of 0.932, ROC-AUC of 0.9940, and PR-AUC of 0.9889. These results validate the effectiveness of both strategies: multi-view integration and 2.5D volumetric representation in enhancing  accuracy. As clearly demostrated, the addition of these ablation in our proposed model yields the best overall performance.
\begin{table}[t!]
\centering
\scriptsize
\caption{Comparison of VLM-based variants across view types and dimensionalities.}
\label{tab:clip_ablation_views}

\setlength{\tabcolsep}{4pt} 

\begin{tabular}{lccccccc}
\toprule
\textbf{VLM Variant} & \textbf{View} & \textbf{Dim.} & \textbf{Prec.} & \textbf{Rec.} & \textbf{F1} & \textbf{ROC} & \textbf{PR} \\
\midrule
Single View & Axial        & 2D   & 0.889 & 0.948 & 0.914 & 0.9921 & 0.9859 \\
 Multi View  & Ax+Cor+Sag   & 2D   & 0.894 & 0.951 & 0.921 & 0.9927 & 0.9868 \\
Single View & Axial        & 2.5D & 0.901 & 0.952 & 0.926 & 0.9933 & 0.9875 \\
\textbf{MOSAIC (Ours)} & Ax+Cor+Sag & \textbf{2.5D} & \textbf{0.934} & \textbf{0.956} & \textbf{0.942} & \textbf{0.9940} & \textbf{0.9889} \\
\bottomrule
\end{tabular}
\end{table}
\begin{table}[b!]
\centering
\caption{Impact of Stage-1 filtering on prediction performance and inference time per volume across models.}
\label{tab:stage1_effect}
\scriptsize
\begin{tabular}{lcccccc}
\toprule
\textbf{Model} & \textbf{Stage-1} & \textbf{Avg Accuracy} & \textbf{Avg F1-score} & \textbf{ROC-AUC} & \textbf{Inference Time (s)} \\
\midrule
EfficientNet-B0     & No  & 0.9939 & 0.9875 & 0.9996 & 4.3  \\
                    & Yes & 0.9914 & 0.9868 & 0.9993 & \textbf{3.2} \\[0.5ex]
MobileNet-V2        & No  & 0.9919 & 0.9845 & 0.9994 & 3.9  \\
                    & Yes & 0.9888 & 0.9813 & 0.9989 & \textbf{2.6} \\[0.5ex]
ResNet-50           & No  & 0.9932 & 0.9867 & 0.9993 & 6.8  \\
                    & Yes & 0.9908 & 0.9856 & 0.9991 & \textbf{5.1} \\[0.5ex]
Swin Transformer-T  & No  & 0.9940 & 0.9881 & 0.9994 & 7.4  \\
                    & Yes & 0.9914 & 0.9865 & 0.9992 & \textbf{5.9} \\[0.5ex]
\textbf{MOSAIC (Ours)}         & No  & 0.9718 & 0.9368 & 0.9968 & 12.4 \\
                    & Yes & 0.9581 & 0.9425 & 0.9940 & \textbf{8.7} \\
\bottomrule
\end{tabular}
\end{table}

\subsubsection{Effect of Stage-1 Filtering}

In Table~\ref{tab:stage1_effect}, we report the average accuracy, F1-score, ROC-AUC, and per-volume inference time for various backbone models, evaluated with and without the proposed Stage-1 filtering module. This can be seen that although our overall pipeline without filtering achieves marginally higher performance in some metrics, typically within 0.2–0.5\%, the addition of Stage-1 significantly reduces inference time, achieving improvements between 20\% and 35\% across all models.This gain is particularly pronounced in transformer-based architectures, where per-slice processing is computationally intensive. Beyond computational efficiency, Stage-1 contributes to improved data balancing by discarding non-informative slices, ensuring that the downstream module focuses on anatomically relevant regions. These results validate the effectiveness of the Stage-1 module as a lightweight and practical pre-processing step that enhances runtime performance while preserving predictive integrity.

\subsubsection{Effect of Slice Localization Concordance (SLC)}
\begin{figure}[t!]
\centering

\begin{subfigure}{0.48\textwidth}
  \centering
  \includegraphics[width=0.79\linewidth]{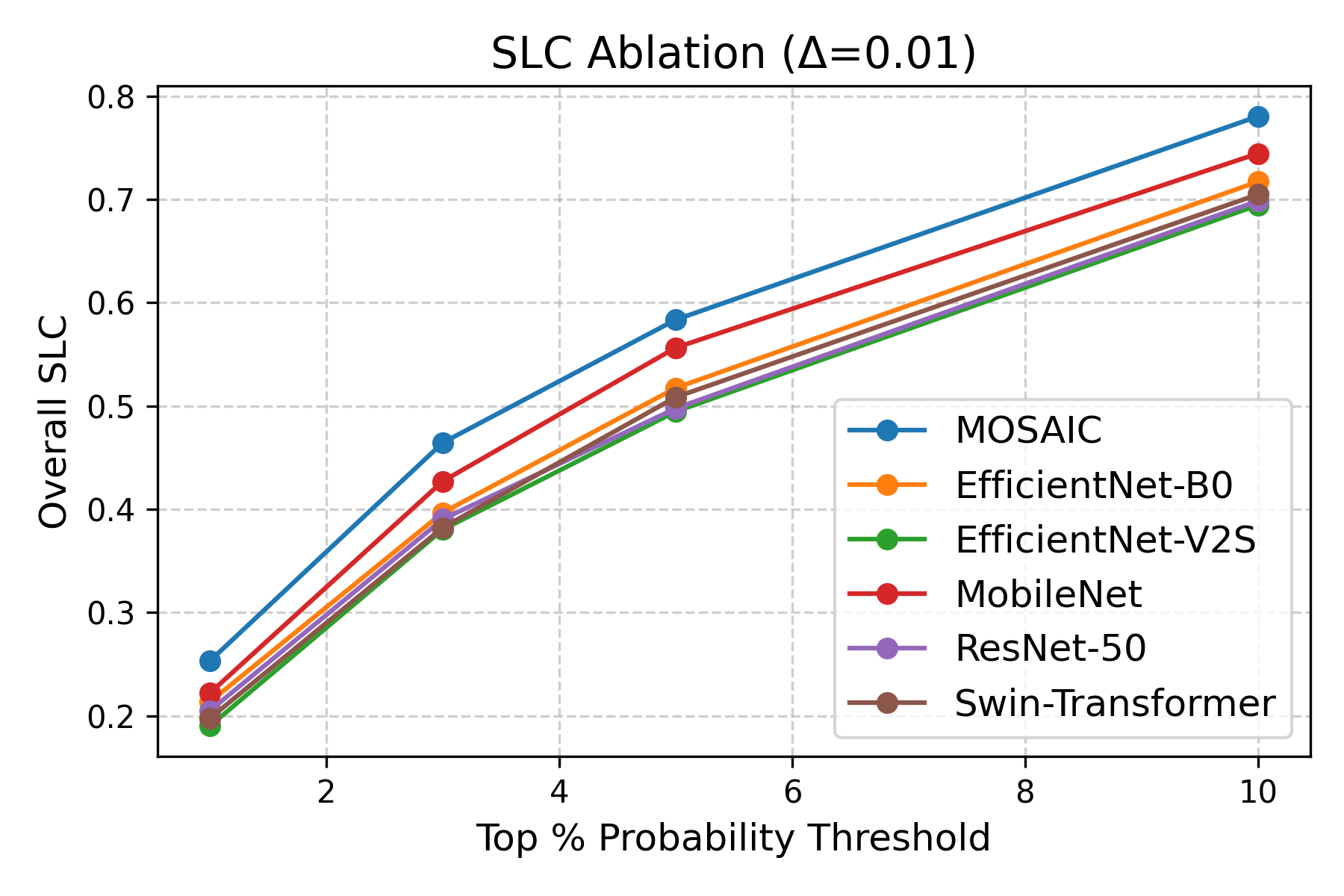}
  \caption{}
\end{subfigure}
\hspace{-5mm}
\begin{subfigure}{0.48\textwidth}
  \centering
  \includegraphics[width=0.79\linewidth]{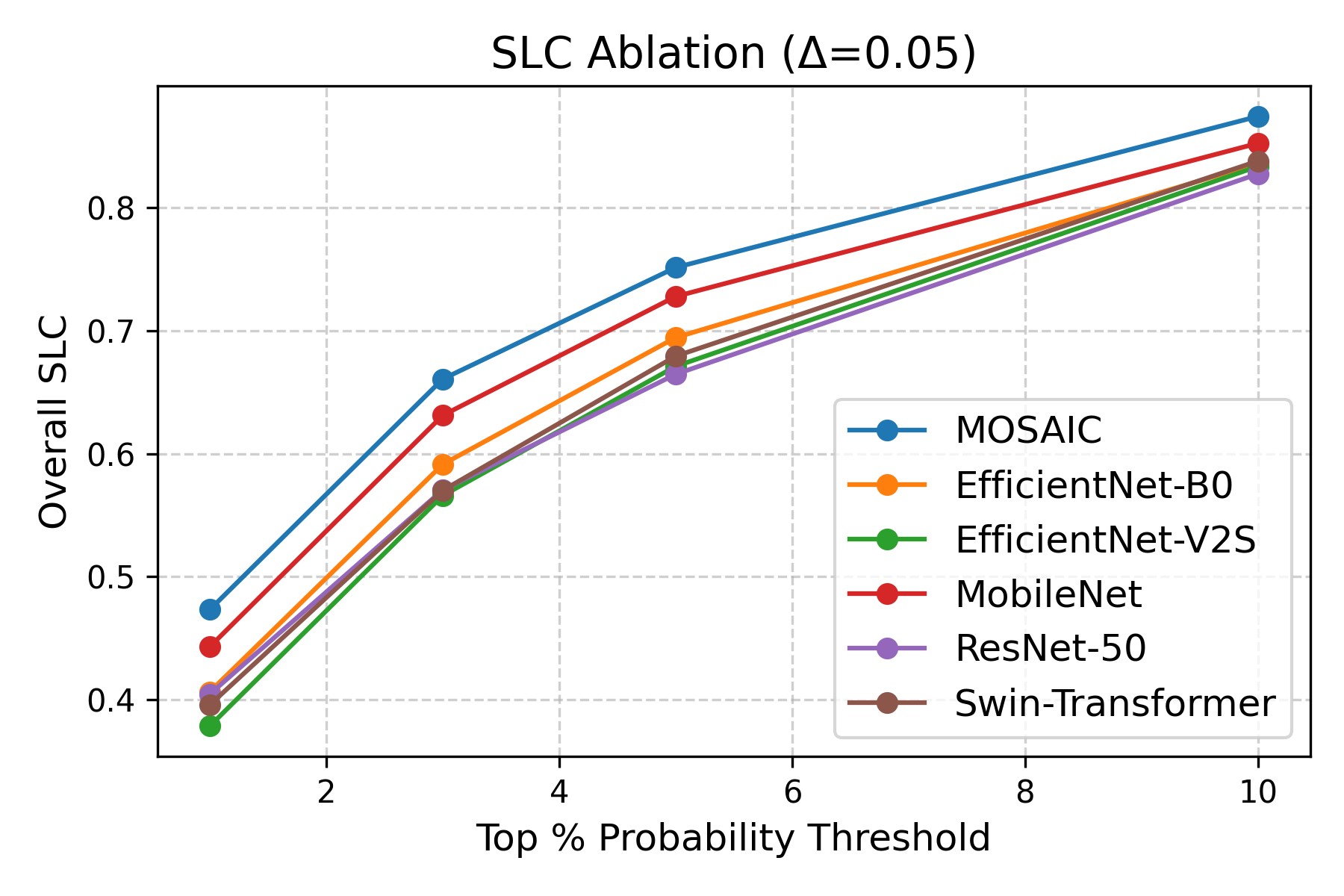}
  \caption{}
\end{subfigure}

\vspace{-1mm}

\begin{subfigure}{0.48\textwidth}
  \centering
  \includegraphics[width=0.79\linewidth]{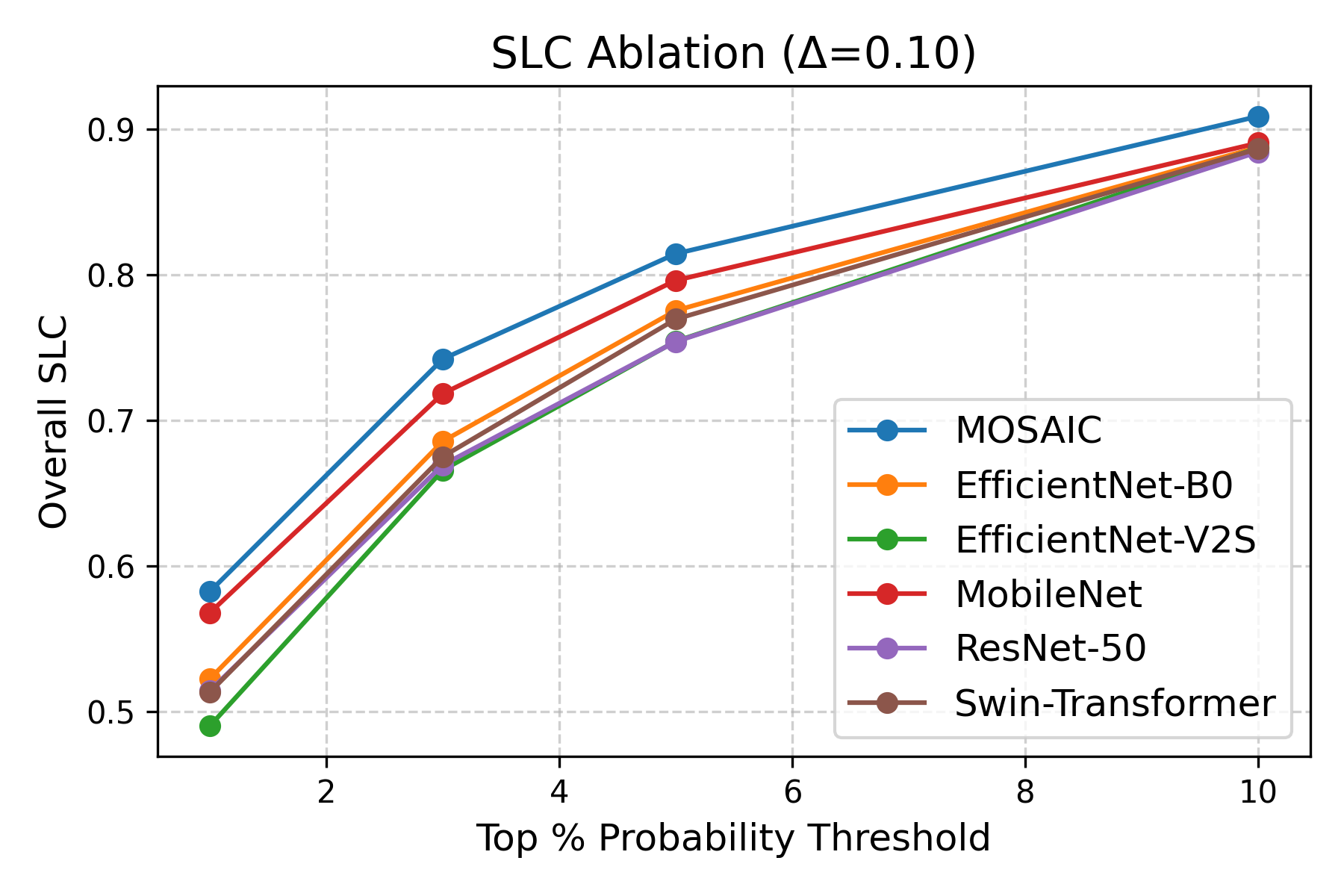}
  \caption{}
\end{subfigure}
\hspace{-5mm}
\begin{subfigure}{0.48\textwidth}
  \centering
  \includegraphics[width=0.79\linewidth]{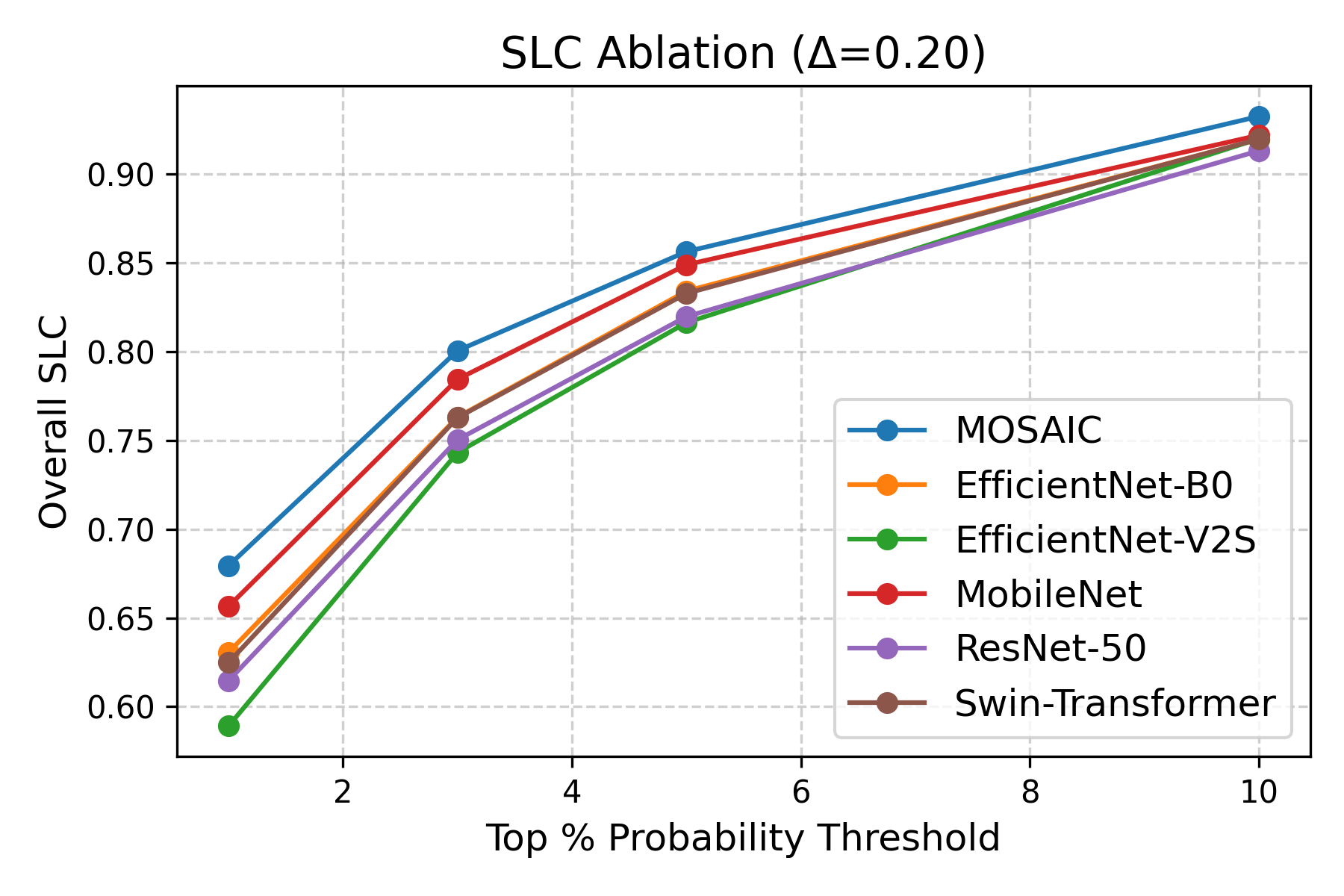}
  \caption{}
\end{subfigure}

\caption{SLC curves at different tolerance thresholds ($\delta$).}
\label{fig:slc_deltas}
\end{figure}

To assess the spatial precision of predicted slices, we evaluate models using Slice Localization Concordance (SLC) across varying tolerance thresholds ($\delta$) and Top-\% selection cutoffs both in Table~\ref{tab:slc-combined} and Figure~\ref{fig:slc_deltas}). We show in Table \ref{tab:slc-combined}, that SLC captures how closely the predicted slice aligns with the anatomically optimal ground truth. Our model consistently outperforms all baselines across every $\delta$ value, especially at lower tolerances (e.g., $\delta = 0.01$ and $0.05$), indicating superior precision in identifying spatially relevant slices. As shown in Figure~\ref{fig:slc_deltas}, the SLC scores for MOSAIC remain robust even as the tolerance range increases, suggesting both high accuracy and generalizability. These results confirm that our proposed model not only ranks slices effectively but does it in a spatially meaningful and anatomically aligned manner which is critical for clinical applicability in organ-specific tasks.

\begin{table*}[b!]
\centering
\scriptsize
\setlength{\tabcolsep}{4pt} 
\renewcommand{\arraystretch}{0.9} 

\caption{Overall Slice Localization Concordance (SLC) for all models across different tolerance values ($\delta$) and Top-\% slice selection thresholds. Higher values indicate better spatial alignment.}
\label{tab:slc-combined}

\begin{tabular}{llcccc}
\toprule
\textbf{Model} & \textbf{$\delta$} & \textbf{Top-1\%} & \textbf{Top-3\%} & \textbf{Top-5\%} & \textbf{Top-10\%} \\
\midrule

\multirow{4}{*}{EfficientNet-B0}
  & 0.01 & 0.1439 & 0.2861 & 0.3928 & 0.5733 \\
  & 0.05 & 0.3470 & 0.4997 & 0.6021 & 0.7384 \\
  & 0.10 & 0.4709 & 0.6169 & 0.6982 & 0.8220 \\
  & 0.20 & 0.5811 & 0.7136 & 0.7825 & 0.8745 \\
\midrule

\multirow{4}{*}{EfficientNet-V2-S}
  & 0.01 & 0.1392 & 0.2839 & 0.3694 & 0.5322 \\
  & 0.05 & 0.2992 & 0.4690 & 0.5661 & 0.7067 \\
  & 0.10 & 0.4124 & 0.5753 & 0.6629 & 0.7845 \\
  & 0.20 & 0.5173 & 0.6692 & 0.7421 & 0.8489 \\
\midrule

\multirow{4}{*}{MobileNet-V2}
  & 0.01 & 0.1416 & 0.2923 & 0.3926 & 0.5860 \\
  & 0.05 & 0.3443 & 0.5120 & 0.6053 & 0.7555 \\
  & 0.10 & 0.4581 & 0.6231 & 0.7149 & 0.8275 \\
  & 0.20 & 0.5717 & 0.7098 & 0.7808 & 0.8751 \\
\midrule

\multirow{4}{*}{ResNet-50}
  & 0.01 & 0.1456 & 0.2857 & 0.3855 & 0.5484 \\
  & 0.05 & 0.3195 & 0.4895 & 0.5862 & 0.7374 \\
  & 0.10 & 0.4348 & 0.5951 & 0.6763 & 0.8117 \\
  & 0.20 & 0.5412 & 0.6890 & 0.7598 & 0.8655 \\
\midrule

\multirow{4}{*}{Swin-T}
  & 0.01 & 0.1514 & 0.2859 & 0.3829 & 0.5629 \\
  & 0.05 & 0.3276 & 0.4767 & 0.5671 & 0.7152 \\
  & 0.10 & 0.4434 & 0.5843 & 0.6638 & 0.7836 \\
  & 0.20 & 0.5510 & 0.6738 & 0.7408 & 0.8444 \\
\midrule

\multirow{4}{*}{\textbf{MOSAIC (Ours)}}
  & 0.01 & \textbf{0.2530} & \textbf{0.4644} & \textbf{0.5834} & \textbf{0.7803} \\
  & 0.05 & \textbf{0.4731} & \textbf{0.6608} & \textbf{0.7515} & \textbf{0.8745} \\
  & 0.10 & \textbf{0.5825} & \textbf{0.7421} & \textbf{0.8144} & \textbf{0.9087} \\
  & 0.20 & \textbf{0.6794} & \textbf{0.8005} & \textbf{0.8563} & \textbf{0.9324} \\
\bottomrule
\end{tabular}
\end{table*}

\section{Results and Discussions}
\subsection{Evaluation of the Stage-1 Filtering Module}

As discussed above in detail, stage-1 aims to discard slices lacking relevant anatomical content across axial, coronal, and sagittal planes. A lightweight binary classifier is trained per view to identify informative slices containing foreground organs. As shown in Table~\ref{tab:performance_metrics}, our model reports high classification performance across all views, with mean F1-scores exceeding 0.97 and ROC-AUC above 0.99, confirming the model’s ability to differentiate informative slices from background slices.

\begin{table}[b!]
\centering
\caption{Performance of Stage-1 filtering across anatomical views.}
\label{tab:performance_metrics}
\begin{tabular}{lcccccc}
\toprule
\textbf{View} & \textbf{Accuracy} & \textbf{Precision} & \textbf{Recall} & \textbf{F1-score} & \textbf{ROC-AUC} & \textbf{Support} \\
\midrule
Axial     & 0.9674 & 0.9639 & 0.9796 & 0.9717 & 0.9954 & 1716 \\
Coronal   & 0.9818 & 0.9931 & 0.9691 & 0.9810 & 0.9986 & 6144 \\
Sagittal  & 0.9771 & 0.9968 & 0.9721 & 0.9843 & 0.9975 & 6144 \\
\bottomrule
\end{tabular}
\end{table} This is worth noting that stage-1 also yields substantial computational benefits. From ~68,000 original slices, only ~40,000 are retained. In Table~\ref{tab:slice_reduction} we show the slice-level statistics where the coronal view exhibits the highest reduction (52.7\%), while sagittal slices are retained more due to spatial consistency. This results in a 41.3\% reduction in the overall computational workload, significantly improving processing efficiency without compromising performance.
\begin{table}[t!]
\centering
\caption{Slice retention statistics across anatomical views post Stage-1 filtering.}
\label{tab:slice_reduction}
\begin{tabular}{lcccc}
\toprule
\textbf{View} & \textbf{Avg. Total Slices} & \textbf{Avg. Retained Slices} & \textbf{Retention Rate} & \textbf{Reduction (\%)} \\
\midrule
Axial     & 143.0  & 83.0    & 0.58 & 41.96 \\
Coronal   & 512.0  & 242.2   & 0.47 & 52.70 \\
Sagittal  & 512.0  & 370.2   & 0.72 & 27.70 \\
\bottomrule
\end{tabular}
\end{table}
More importantly, the filtering step especially benefits organs that are spatially dispersed or occur intermittently, such as, the pancreas and aorta, by eliminating noisy, non-informative slices and improving the input to Stage-2 focus. This is worth mentioning that low-contrast regions like the gallbladder also see performance gains due to reduced background interference.



\subsection{Evaluation of the Multi-View Slice Selector (MOSAIC)}

The proposed MOSAIC (Multi-view Organ Slice Selector for Anatomically Informed CT localization) framework demonstrates strong per-organ prediction and localization capabilities across the BTCV dataset. In Table~\ref{tab:cross_model_metrics}, we presents the organ-wise quantitative metrics. Notably, high F1-scores are observed across most organs, including anatomically large structures such as the liver, spleen, and kidneys, but also for smaller or lower-contrast structures like the adrenal glands and gallbladder. For instance, the right adrenal achieves an F1-score of 0.846 and PR-AUC of 0.947, underscoring the model’s ability to reliably detect anatomically challenging regions. The high ROC-AUC and PR-AUC values (frequently exceeding 0.98) further affirm the model’s discriminative capability and robustness across organ classes.

\begin{table}[b!]
\centering
\scriptsize
\caption{Performance metrics of the proposed MOSAIC model per organ on BTCV.}
\label{tab:cross_model_metrics}
\begin{tabular}{lccccc}
\toprule
\textbf{Organ} & \textbf{Precision} & \textbf{Recall} & \textbf{F1-score} & \textbf{ROC-AUC} & \textbf{PR-AUC} \\
\midrule
Spleen & 0.946 & 0.962 & 0.954 & 0.992 & 0.989 \\
Right Kidney & 0.919 & 0.952 & 0.935 & 0.990 & 0.983 \\
Left Kidney & 0.910 & 0.961 & 0.934 & 0.990 & 0.981 \\
Gallbladder & 0.745 & 0.975 & 0.845 & 0.988 & 0.940 \\
Esophagus & 0.838 & 0.979 & 0.903 & 0.994 & 0.972 \\
Liver & 0.994 & 0.923 & 0.957 & 0.984 & 0.997 \\
Stomach & 0.958 & 0.922 & 0.940 & 0.984 & 0.989 \\
Aorta & 0.929 & 0.963 & 0.946 & 0.993 & 0.988 \\
Inferior Vena Cava & 0.901 & 0.971 & 0.935 & 0.993 & 0.986 \\
Portal Vein & 0.970 & 0.926 & 0.948 & 0.987 & 0.993 \\
Pancreas & 0.950 & 0.927 & 0.938 & 0.986 & 0.988 \\
Right Adrenal & 0.742 & 0.983 & 0.846 & 0.991 & 0.947 \\
Left Adrenal & 0.766 & 0.974 & 0.858 & 0.989 & 0.942 \\
\bottomrule
\end{tabular}
\end{table}

\subsection{Multi-View SLC Evaluation}

\begin{figure}[t!]
    \centering
    \vspace{-1em}
    \includegraphics[width=\linewidth]{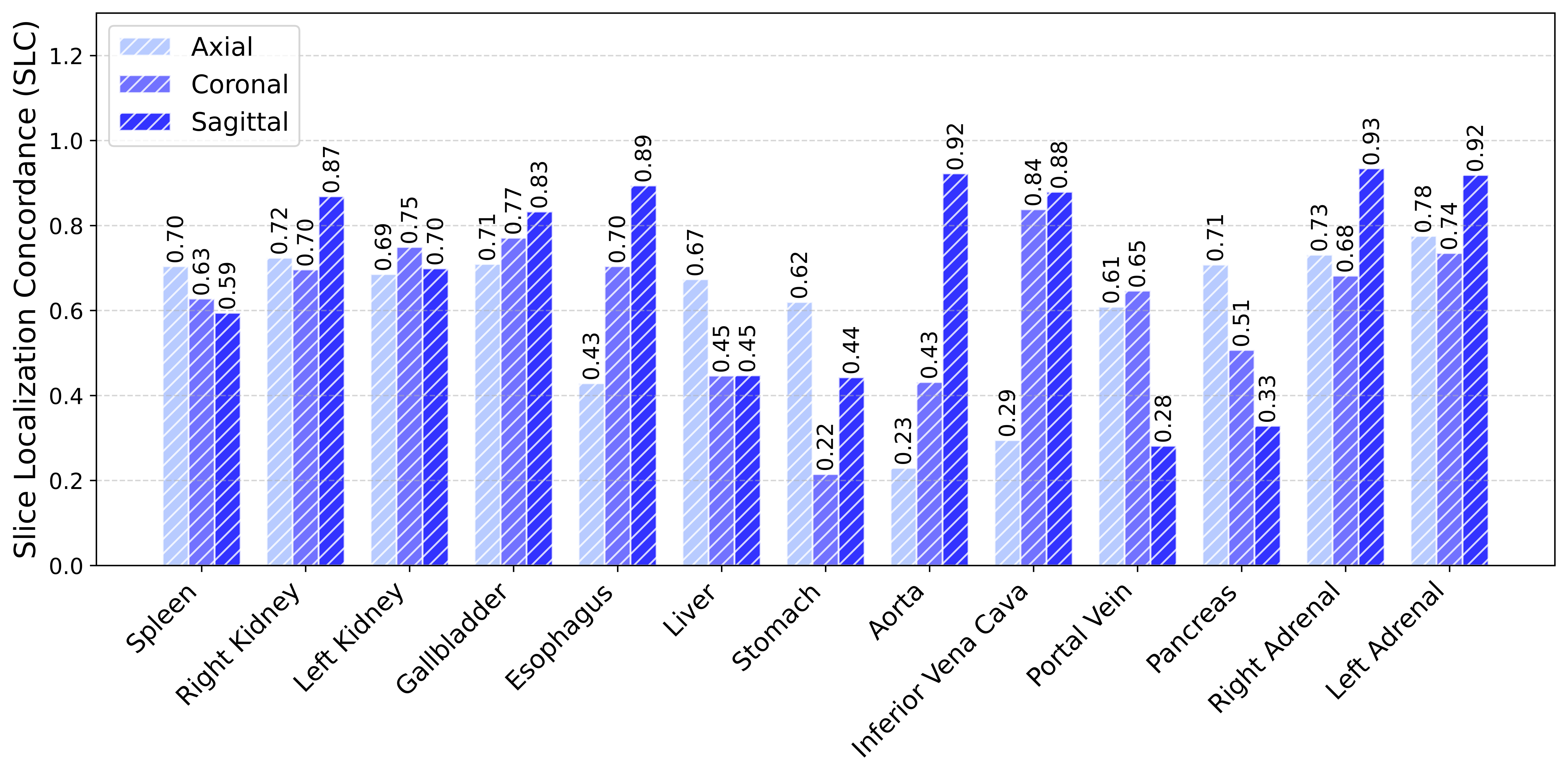}
    \vspace{-1em}
    \caption{Per-organ mean Slice Localization Concordance (SLC) across axial, coronal, and sagittal views.}
    \label{fig:slc_views}
    \vspace{-0.5em}
\end{figure}

To assess spatial localization performance across anatomical orientations, we report mean Slice Localization Concordance (SLC) scores for each organ across axial, coronal, and sagittal views, as shown in Figure~\ref{fig:slc_views}.This can be noted that SLC serves as a critical metric in answering an often-overlooked question: \textit{which anatomical view is best suited for localizing a specific organ?} By comparing the predicted slice locations to ground truth anatomical positions, SLC quantifies the spatial alignment quality across views, offering actionable insights for model design. The results reveal view-specific strengths that directly inform downstream tasks like segmentation. Axial slices consistently achieve higher SLC scores for large, horizontally extensive organs such as the liver, pancreas, and stomach. In contrast the sagittal views, show superior alignment for vertically elongated or smaller structures like the adrenal glands, esophagus, and the aorta. Similarly, coronal slices offer competitive performance for midline organs including the portal vein, splenic vein, and kidneys. These findings validate the utility of incorporating all three orthogonal views in our 2.5D slice selection pipeline. Moreover, SLC provides a principled basis for view prioritization in hybrid segmentation frameworks enabling selective processing of views that offer maximal anatomical context for each target organ.

\subsection{Qualitative Visualization of Slice Selection Explainability}

\begin{figure}[t!]
\centering

\begin{subfigure}{0.48\textwidth}
  \centering
  \includegraphics[width=\linewidth]{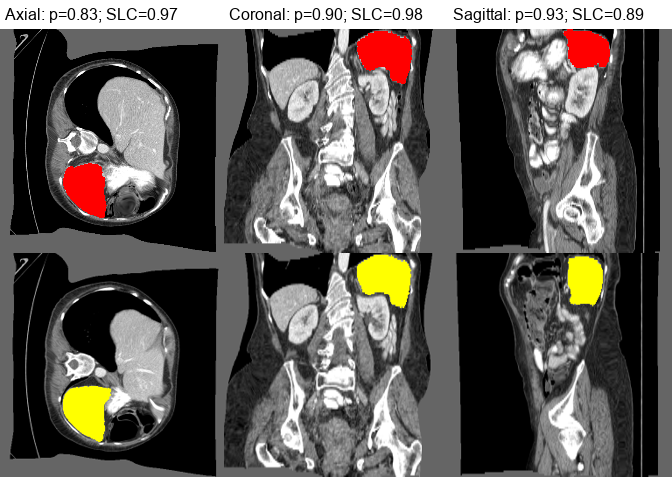}
  \caption{}
\end{subfigure}
\hfill
\begin{subfigure}{0.48\textwidth}
  \centering
  \includegraphics[width=\linewidth]{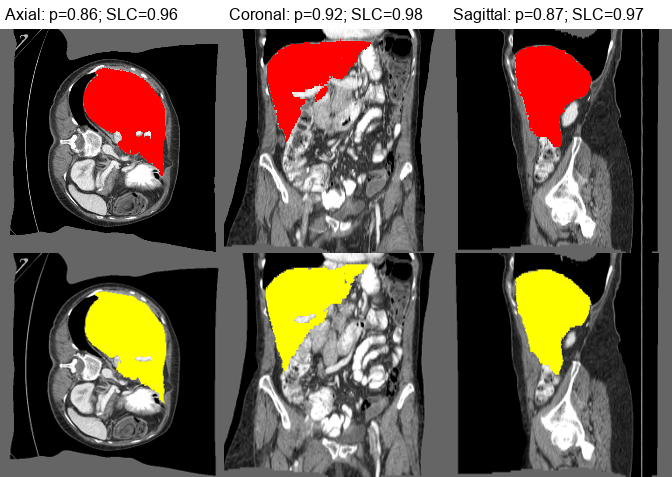}
  \caption{}
\end{subfigure}

\vspace{-0.8mm}

\begin{subfigure}{0.48\textwidth}
  \centering
  \includegraphics[width=\linewidth]{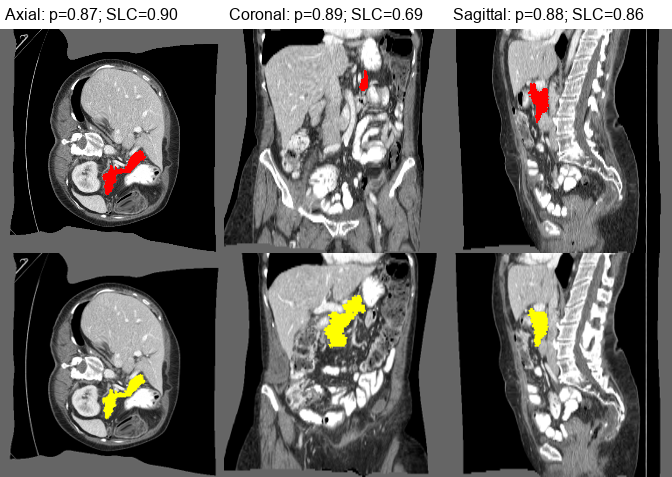}
  \caption{}
\end{subfigure}
\hfill
\begin{subfigure}{0.48\textwidth}
  \centering
  \includegraphics[width=\linewidth]{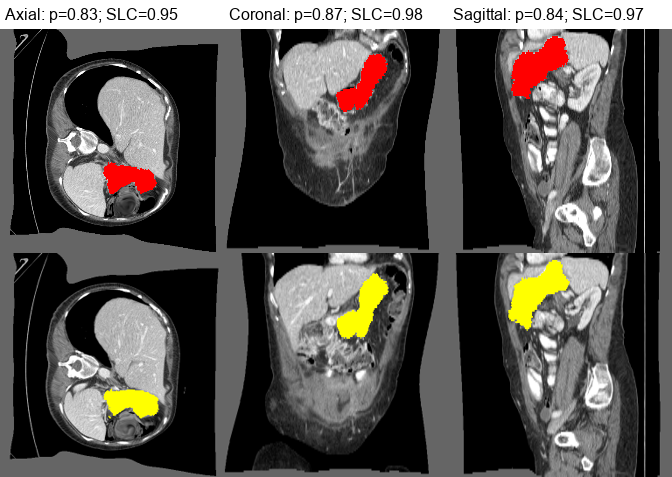}
  \caption{}
\end{subfigure}

\vspace{-0.8mm}

\begin{subfigure}{0.48\textwidth}
  \centering
  \includegraphics[width=\linewidth]{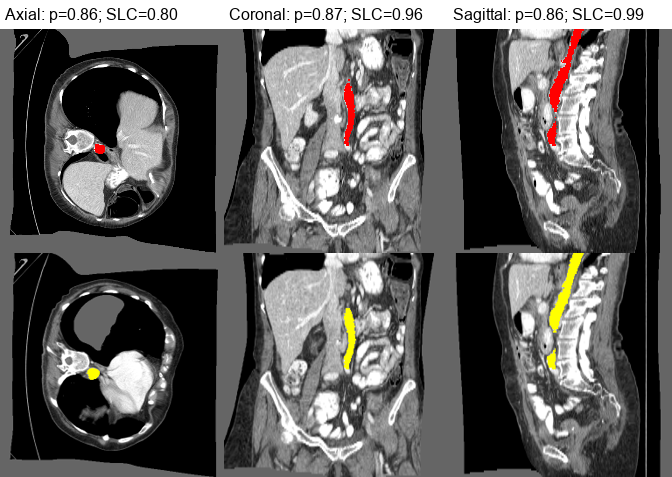}
  \caption{}
\end{subfigure}
\hfill
\begin{subfigure}{0.48\textwidth}
  \centering
  \includegraphics[width=\linewidth]{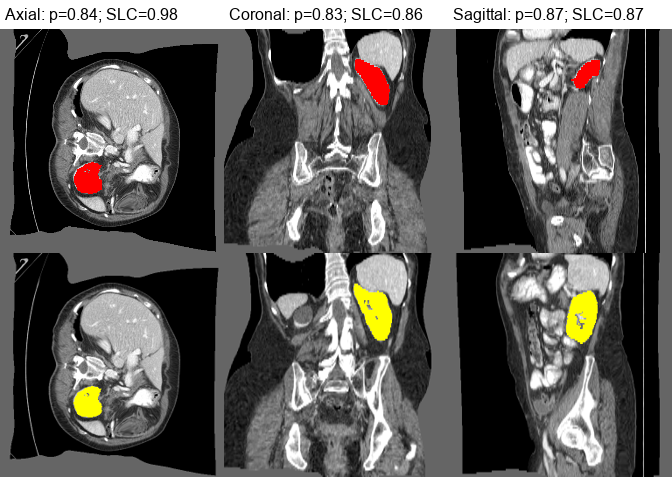}
  \caption{}
\end{subfigure}
\caption{Qualitative examples of high-confidence, high-accuracy slice selection for (a) spleen, (b) liver, (c) pancreas, (d) stomach, (e) aorta, and (f) left kidney. Each panel includes predicted (red) and ground-truth (yellow) slices across axial, coronal, and sagittal views. Prediction confidence (p) and localization agreement (SLC) are shown per view.}
\label{fig:best_slices}
\end{figure}

To further interpret the model’s localization quality, we present qualitative visualizations of both high-performing and failure cases. Figure~\ref{fig:best_slices} shows model-selected slices for six representative organs, each visualized across axial, coronal, and sagittal planes. For every slice, we overlay the model’s predicted peak-probability slice (in red) and the ground-truth anatomical slice (in yellow), annotated with both the prediction confidence (p-score) and the corresponding SLC score. Notably, for large and spatially consistent organs like the liver and spleen, all three views exhibit high spatial alignment. For instance, in the spleen, the axial slice with a p-score of 0.83 achieves an SLC of 0.97, indicating near-perfect localization. Similarly, the pancreas axial slice shows p = 0.87 and SLC = 0.90, and the liver (coronal) reaches a strong p = 0.92 with SLC = 0.98. These trends demonstrate the robustness of the model across views for high-contrast, voluminous structures.

\begin{figure}[t!]
\centering

\begin{subfigure}{0.48\textwidth}
  \centering
  \includegraphics[width=\linewidth]{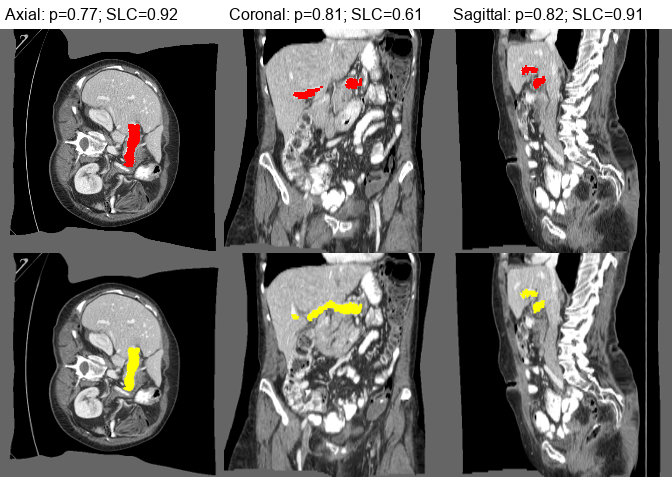}
  \caption{}
\end{subfigure}
\hfill
\begin{subfigure}{0.48\textwidth}
  \centering
  \includegraphics[width=\linewidth]{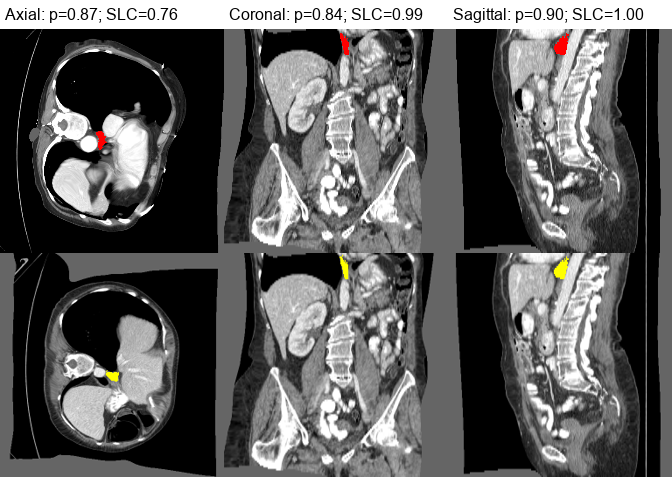}
  \caption{}
\end{subfigure}

\vspace{-1.8mm}

\begin{subfigure}{0.48\textwidth}
  \centering
  \includegraphics[width=\linewidth]{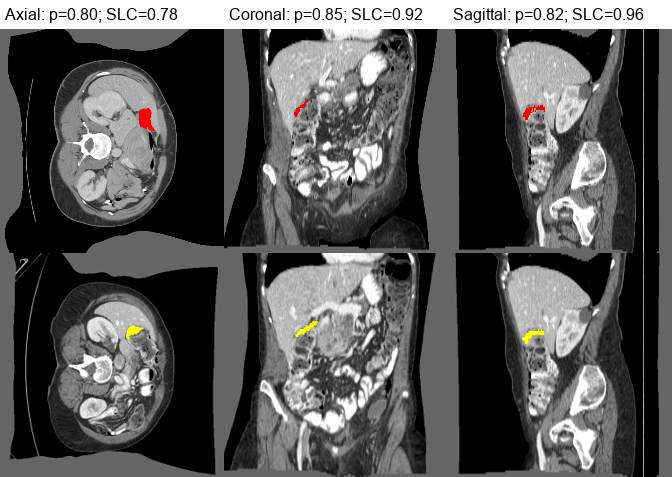}
  \caption{}
\end{subfigure}
\hfill
\begin{subfigure}{0.48\textwidth}
  \centering
  \includegraphics[width=\linewidth]{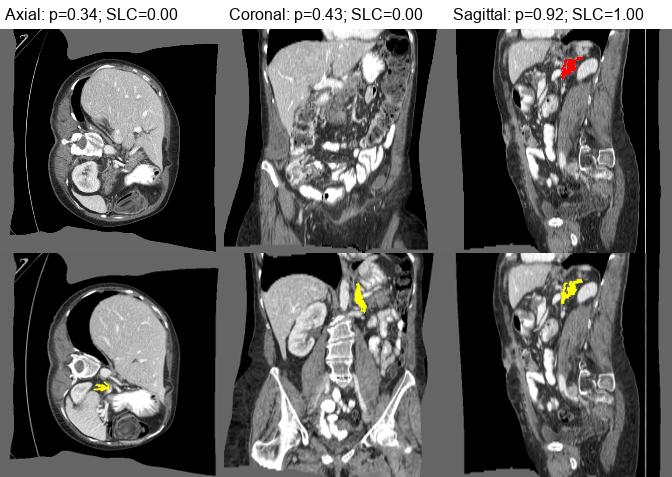}
  \caption{}
\end{subfigure}
\caption{Challenging cases involving small or low-contrast organs: (a) portal vein, (b) esophagus, (c) gallbladder, and (d) left adrenal gland.}
\label{fig:weak_slices}
\end{figure}

More importanty, we present challenging cases in Figure~\ref{fig:weak_slices} involving smaller and lower-contrast organs such as the portal vein, esophagus, gallbladder, and left adrenal gland. These examples highlight the difficulty of achieving accurate localization when certain views contain little to no anatomical evidence. For the left adrenal gland, both axial (p = 0.34) and coronal (p = 0.43) views result in incorrect predictions, reflected in SLC scores of 0.0. However, the sagittal view achieves an almost perfect localization (p = 0.92, SLC = 0.99), indicating that the model learns to rely on the most informative orientation. A similar pattern is observed for the esophagus, where despite modest prediction confidence scores (p = 0.87 axial, 0.84 coronal), all SLCs remain high (> 0.99), showcasing the model's precision even under visual ambiguity. In the case of the gallbladder, a relatively low axial prediction confidence score of 0.80 corresponds to an SLC of only 0.78, while sagittal SLC improves to 0.95. For the portal vein, although all prediction confidence scores remain moderate (~0.77–0.82), the coronal SLC is notably poor (0.61), whereas axial and sagittal slices align well (SLCs = 0.91 and 0.91). These examples confirm the necessity of our multi-view approach, particularly for small or ambiguous organs where localization in one or more views may fail. The model learns to compensate by leveraging the most reliable anatomical perspective, thereby maintaining accurate performance where conventional confidence-based metrics would become insufficient.

\section{Conclusion}

We presented MOSAIC: a multi-view, anatomically-aware 2.5D slice selection framework that integrates vision–language supervision with cross-orientation fusion for efficient organ localization in abdominal CT volumes. Our method introduces a lightweight relevance filtering stage and a VLM-based multi-organ slice selector operating over fused axial, coronal, and sagittal slices. To assess anatomical alignment, we introduced a novel metric designed to measure slice-level localization fidelity with organ-level precision. This metric complements standard classification scores and offers a more targeted evaluation of diagnostic slice quality. Through extensive evaluation on the BTCV benchmark, MOSAIC achieved strong generalization across 13 organs, with a mean F1-score of 0.94, ROC-AUC of 0.994, PR-AUC of 0.993, and an Slice Localization Concordance (SLC) of 0.956, outperforming established 2D and transformer-based backbones. Importantly, our framework offers substantial computational benefits by eliminating redundant slices early in the pipeline, reducing inference cost without sacrificing spatial fidelity. Visual and quantitative results demonstrate that MOSAIC maintains strong localization even for small and challenging structures such as the pancreas, adrenal glands, and esophagus. This study aims to further advance anatomically-aware and language-driven representation learning in medical imaging.




\section*{Data Availability}

The abdominal CT scans used in this study are from the \textit{Beyond the Cranial Vault (BTCV)} dataset, originally released as part of the \textit{Multi-Atlas Labeling Beyond the Cranial Vault} challenge. The dataset is publicly available and can be accessed through the Synapse platform at:\url{https://www.synapse.org/#!Synapse:syn3193805}

Access requires registration and acceptance of the Synapse Data Use Agreement. All usage in this study complies with the associated licensing terms.

\bibliographystyle{plainnat}
\bibliography{cas-refs}

\end{document}